\documentclass[pre,aps,twocolumn]{revtex4-1}
\usepackage{graphicx} 
\usepackage[usenames]{color}
\usepackage{amsmath,amssymb}
\usepackage{subfigure}
\usepackage{gensymb}
\usepackage{natbib}
\usepackage{color}
\usepackage{longtable}
\usepackage{booktabs}
\usepackage{pbox}
\def\strutdepth{\dp\strutbox}
\def\nw#1{\strut\vadjust{\kern-\strutdepth\vtop to0pt{\vss\hbox to\hsize
{\hskip\hsize\hskip5pt$\leftarrow$\hss\strut}}}{\em #1}}

\begin{document}

\title{Formation and post-formation dynamics of bacterial biofilm streamers as highly viscous liquid jets}
\author{Siddhartha Das}\email{sdas1@ualberta.ca}
\author{Aloke Kumar}
\affiliation{Department of Mechanical Engineering, University of Alberta, Edmonton, Alberta, Canada T6G 2G8}

\date{\today}%

\begin{abstract}
It has been recently reported that in presence of low Reynolds number ($Re\ll1$) transport, preformed bacterial biofilms, several hours after their formation, may degenerate in form of filamentous structures, known as streamers. In this letter, we explain that such streamers form as the highly viscous liquid states of the intrinsically viscoelastic biofilms. Such ``viscous liquid" state can be hypothesized by noting that the time of appearance of the streamers is substantially larger than the viscoelastic relaxation time scale of the biofilms, and this appearance is explained by the inability of a viscous liquid to withstand an external shear. Further, by identifying the post formation dynamics of the streamers as that of a viscous liquid jet in a surrounding flow field, we can interpret several unexplained issues associated with the post-formation dynamics of streamers, such as the clogging of the flow passage or the exponential time growth of streamer dimensions.
\end{abstract}

\maketitle

The normal form of bacterial growth in most environments is now recognized to occur as a biofilm, which is a social form of growth associated with solid phase surfaces \cite{CosARM1995,HauJB2010,NeeChap2012,WongMRSB2011}. Biofilms include differentiated populations of cells embedded in a matrix of self-produced extracellular polymeric substances (EPS) \cite{FleNRM2010,WikMRSB2011}, displaying physiological properties that vary significantly from that of a dispersed cell population. Biofilms have attracted significant interdisciplinary attention as they can lead to persistent infections \cite{CosSCI1999,FuxTM2005}, fouling of surfaces \cite{CalNC2011,LapBF1989}, and at the same time help in waste-water treatment \cite{VanTB1993}. 

Biofilms are excellent examples of viscoelastic materials \cite{KlaBB2002,ShaPRL2004} exhibiting a complex range of behaviours to external force including deformation, fracture and strain-hardening \cite{WikMRSB2011}.
Recently, multiple researchers have demonstrated that even in low Reynolds number ($Re\ll1$) flows, appearance of surface-hugging biofilms was followed, after a time lag of several hours, by the appearance of filamentous structures (extruding from the pre-formed biofilms) known as streamers \cite{MarBF2012,RusJRSI2010,RusBPJ2011,ValLOC2012,YazBMF2012,WeaAEM2012}. These streamers that we study are in creeping ($Re\ll1$) background flow (e.g., $Re\sim0.1$ in Rusconi et al. \cite{RusBPJ2011} and Drescher et al. \cite{DrePNAS2013} and $Re\sim0.01$ in Valiei et al. \cite{ValLOC2012}) and hence are distinctly different from the streamers formed in turbulent background flow in a multitude of scenarios \cite{LewWST1995,StoBB1998,StoJAM1999,HalNRM2004,MeyEM2011,HalAEM2006,StoJIMB2002} (see Ref. \cite{Turb_Streamer} for more details). Streamer formation in low $Re$ has wide repercussions as they can act as precursors to the formation of mature biofilms in complex microstructures \cite{ValLOC2012,DrePNAS2013}, lead to more rapid and catastrophic clogging of devices \cite{DrePNAS2013}, cause substantial flow-structure interactions \cite{TahBPJ2012}, etc. Despite the recent interests in biofilm streamer dynamics, there remain several open questions, e.g., What is the effect of biofilm rheology in streamer formation?  What is the reason for the substantial time lag between the formation of biofilms and the appearance of streamers? How can one explain different effects associated with the post-formation dynamics of streamers, such as the rapid clogging of the flow device \cite{DrePNAS2013}, or the very fast growth of the streamer dimensions with time \cite{RusBPJ2011}? 

In this letter, we provide answers to all of the above questions. We start by explaining that the streamers form as the ``\textit{viscous liquid}" state of the intrinsically viscoelastic biofilms (with shear modulus $G$, viscosity $\mu_b$ and the viscoelastic relaxation time $\tau_{ve}=\mu_b/G$). Such a hypothesis allows us to explain the large time lag (henceforth denoted as $t_s$) between the formation of the biofilms and the appearance of the streamers, and at the same time quantify the role of biofilm rheology in streamer formation. Being in the ``viscous liquid" state, the biofilms fail to resist the flow-driven shear forces (often too weak to cause any substantial elastic extrusion) resulting in degeneration as streamers. Secondly, we explain the post-formation dynamics of the streamers as that of a highly viscous liquid jet in a background flow. In case the background flow can be approximated to be co-axial to the streamer jet transport, we demonstrate that the typical conditions pertaining to streamer formation \cite{RusJRSI2010,RusBPJ2011,ValLOC2012,WeaAEM2012,DrePNAS2013} will lead to ``absolute instability" \cite{GuiPRL2007,GuiPRE2008} of the streamer liquid jet, enforcing the jet to break down into smaller drops. We derive scaling relationships to quantify the breakup length characterizing such breakup, and demonstrate that these lengths are often too large to cause any drop formation in microfluidic systems studying streamer formation \cite{ValLOC2012}. On the contrary, dictated by the geometry, if the streamer jet is in ``crossflow" to the background flow \cite{DrePNAS2013}, the streamers breakdown into drops almost instantly after their formation. This can explain the unbroken filamentous morphology of streamers in the experiment of Valiei et al. \cite{ValLOC2012}, and at the same time account for the ``porous-matrix-like" structure inside the flow domain witnessed in the experiment of Drescher et al. \cite{DrePNAS2013}. Finally, we establish that only by  considering the streamers to evolve as viscous drops, we can quantify effects such as exponential increase in streamer dimensions \cite{RusBPJ2011,DrePNAS2013} and catastrophic clogging of flow device. 

%
%
\begin{figure}[t!]
\centering
\includegraphics[width=6.5cm, height=3.3cm]{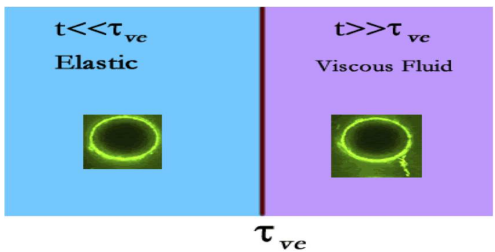}
\caption{(Color Online) Pictorial representation of the proposed hypothesis that streamers form as viscous liquid jets. On the left hand side, we show green flourescent bacteria forming a biofilm around a 50 $\mu m$ diameter micropost (experimental set up identical to that of Valiei et al. \cite{ValLOC2012}) at a time $t$, such that $t\ll\tau_{ve}$ -- hence the biofilm behaves as an elastic solid. On the right hand side, we show that the same biofilm is degenerating into a streamer as a viscous liquid jet at a time $t$ such that $t\gg\tau_{ve}$.}
\label{fig:schem}       
\end{figure}  

\textit{Streamers form as ``viscous liquid" state of biofilms}: Biofilms are viscoelastic liquids \cite{KlaBB2002,ShaPRL2004,StoJIMB2002,WinPRSA2011} -- therefore, at times $t\ll\tau_{ve}$ they exhibit a behaviour analogous to elastic solids, whereas at times $t\gg\tau_{ve}$ they exhibit a behaviour identical to that of highly viscous liquids. Rheological measurements exhibit wide ranges of values of the shear modulus $G$ and viscosity $\mu_b$ of the biofilms \cite{KlaBB2002,ShaPRL2004,StoJIMB2002,StoBB1999,TowBF2003,HouBF2008,SteMB2009,LauBPJ2009,HohLANG2009}, although there is a remarkable commonality in the viscoelastic relaxation time $\tau_{ve}=\mu_b/G$ \cite{ShaPRL2004}. In order to understand the rheological state of the biofilms that lead to streamer formation, we must compare $t_s$ with $\tau_{ve}$. In Table \ref{tab:data}, we summarize the $t_s$ values corresponding to different experiments reporting the formation of streamers, as well as the corresponding $\tau_{ve}$ values (for the biofilms forming the streamers) obtained from separate rheological measurements. From this table it is clear that we always encounter $t_s\gg\tau_{ve}$, establishing the validity of our hypothesis of considering streamers as ``viscous liquid" state of the biofilms (see Fig.~\ref{fig:schem} and Ref. \cite{Supp} for more details). Please note that in an earlier study, Rusconi et al. \cite{RusJRSI2010} used this idea of $t_s\gg\tau_{ve}$ to hypothesize streamers as viscous liquid; however, they did not provide any further analysis to establish their claims. Also we shall like to distinguish between the streamers that we describe from that of the aggregation-driven streamers witnessed by Yazdi and Ardekani \cite{YazBMF2012}.
\begin{table}[htbp]
  \centering
  \caption{Variation of the time scales $\tau_{ve}$ and $t_s$ from the corresponding biofilms}
    \begin{tabular}{|c|c|c|}
    \toprule
    \hline
    \textbf{Bacteria forming Biofilms} & \textbf{$\tau_{ve}$ (min)} & \textbf{$t_s$(hours)}  \\
    \midrule
     \hline
    {\textit{Pseudomonas aeruginosa}} & 18 \cite{ShaPRL2004}   & 5--10 \cite{RusJRSI2010,RusBPJ2011}, \newline\\
    & & 20--40 \cite{DrePNAS2013} \\
     \hline
     {\textit{Pseudomonas fluorescens}} & 18 \cite{ShaPRL2004,note_Pse_Flor}   & 9 \cite{ValLOC2012} \\
     \hline
    {\textit{Staphylococcus epidermis}} & \pbox{100cm}{19.2 \cite{SteMB2009}, \\21.9--25.5 \cite{IanIJMS2011}}  & 6 \cite{WeaAEM2012} \\
     \hline
    \bottomrule
    \end{tabular}%
  \label{tab:data}%
\end{table}%
Our analysis negates the idea that the streamers form from the ``elastic" degeneration of the biofilms \cite{RusBPJ2011,AutPOF2011,DrePNAS2013}, since the imposed elastic strain ($e$) from the flow shear is invariably very weak, i.e., $e\sim10^{-2}-10^{-4}$ \cite{RusJRSI2010,RusBPJ2011,ValLOC2012,WeaAEM2012,DrePNAS2013} (see Table II in \cite{Supp} for details). On the contrary, when the biofilms attain the ``viscous liquid" state, it will fail to resist any imposed shear, thereby degenerating into streamers (streamers are therefore jets of viscous liquids, see later). 
Note that the quantitative relationship between the strain $e$ (or applied stress $\sigma$) and $t_s$, obtained from different experiments, are not well explained (see Ref. \cite{Stress_timescale} for more details).  

\begin{figure}[t!]
\centering
\includegraphics[width=9cm, height=4.6cm]{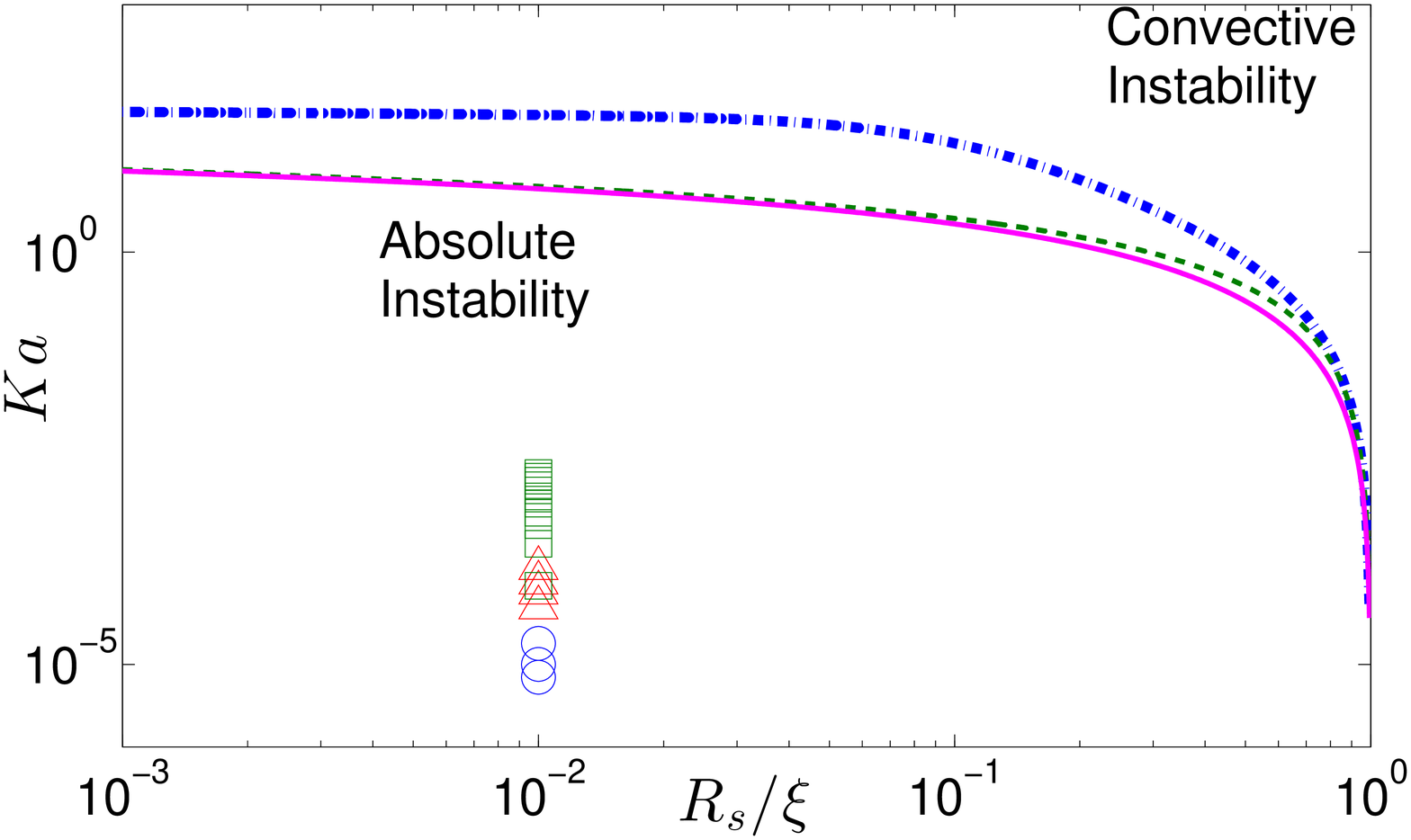}
\caption{(Color Online) Phase diagram for the instability (for a circular jet in a circular capillary) in the $Ka=(-\partial p/\partial z)\xi^2/\gamma=\mu_fu_c/\gamma$ and $R_s/\xi$ planes for different $\mu_b/\mu_f$ values (blue dashdot line for $\mu_b/\mu_f=0.01$, green dashed line for $\mu_b/\mu_f=1$ and magenta solid line for $\mu_b/\mu_f\ge100$). We have convective instability above the curves and absolute instability below the curves. Absolute instability leads to spontaneous breakdown of the jet into drops. We plot the experimental results [i.e., the corresponding $Ka$ (see Table III in \cite{Supp}) and $R_s/\xi$ or $R_s/(h/2)$ values; here $h$ is the characteristic dimension of a possible rectangular geometry, see \cite{Supp}] for different experiments (circle for Valiei et al. \cite{ValLOC2012}, triangles for Rusconi et al. \cite{RusBPJ2011} and squares for Drescher et al. \cite{DrePNAS2013}). Also for all the experiments $\mu_b/\mu_f=10^7$ (since we take $\mu_f=10^{-3}~Pa-s$ and $\mu_b=10^{4}~Pa-s$ \cite{ShaPRL2004}) and we take $R_s/\xi~[or~R_s/(h/2)]=0.01$. Therefore the experimental $Ka$ and $R_s/\xi$ values signify an absolute instability regime, which will suggest a spontaneous breakdown of the jet into droplets for all the streamer-forming experiments \cite{ValLOC2012,RusBPJ2011,DrePNAS2013}.}
\label{fig:Stab}       
\end{figure}  

\textit{Post-formation streamer dynamics -- streamers as highly viscous liquid jets}: Streamers being formed from the ``viscous liquid" biofilms, their post-formation dynamics can be interpreted as that of high viscous liquid jets moving through a background flow. This dynamics depends on the direction of the background flow with respect to the direction of the streamer jet. In case the background flow is in the same direction as that of the streamers, we can invoke the study of Guillot et al. \cite{GuiPRL2007,GuiPRE2008} to describe the streamer dynamics: we perform stability analysis of the streamers, represented by a thin cylindrical viscous jet of viscosity $\mu_b$ and radius $R_s$, moving coaxially with a flow (of viscosity $\mu_f$) inside a capillary of radius $\xi$ (see \cite{Supp} and Fig. 1 in Ref. \cite{Supp}). 
The instability phase diagram, characterized by the parameters $Ka=(-\partial p/\partial z)\xi^2/\gamma=\mu_fu_c/\gamma$ ($\partial p/\partial z$ is the pressure gradient, $u_c$ is the characteristic speed of the surrounding liquid, $\gamma$ is the surface tension between the streamer liquid and the bulk liquid; $Ka$ is an effective capillary number at the scale of the capillary) and ratios $R_s/\xi$ and $\mu_b/\mu_f$, is shown in Fig.~\ref{fig:Stab} (see \cite{Supp} for analytical expressions characterizing the instability behaviour). Above the lines the viscous streamer jet will be convectively unstable, whereas below the lines the jet is absolutely unstable \cite{GuiPRL2007,GuiPRE2008}. 
The ``absolute instability" regime is characterized by spontaneous breakdown of the jet into drops with the perturbations propagating backwards. In Fig.~\ref{fig:Stab}, we plot the experimental conditions (characterized by $Ka$ and $R_s/\xi$ values; see Table III in \cite{Supp} for determination of these parameters) corresponding to the streamer formation, as reported in the literature \cite{RusBPJ2011,ValLOC2012,DrePNAS2013}. For these experiments, we typically encounter $\mu_b/\mu_f\sim10^7$, and accordingly these data points are located below the instability phase line, indicating that the streamer formation conditions are such that the streamer liquid jet will spontaneously break down into smaller drops, with the breakdown being characterized by the break up length $\ell_{bs}$. Following Javadi et al. \cite{JavPRL2013}, we can develop a scaling argument to quantify this $\ell_{bs}$ for streamers in coaxial flow, as $\ell_{bs}\sim R_s(\mu_bu_c/\gamma)$ (see \cite{Supp}). Consequently, for the experiment of Valiei et al. \cite{ValLOC2012}, we get $\ell_{bs}\sim100~\mu m$ \cite{Supp} --  we indeed find that the streamers continue as long unbroken jets/filaments between two microposts having separation distance much smaller than $\ell_{bs}$ (see Fig. 2 in Ref. \cite{Supp}). Using the same scaling, we would get  $\ell_{bs}\sim1-10~mm$ \cite{Supp} for the experiment of Rusconi et al. \cite{RusBPJ2011} and Drescher et al. \cite{DrePNAS2013}. However in these experiments \cite{RusBPJ2011,DrePNAS2013}, as per our hypothesis,  the streamer viscous jet will rapidly break down into smaller drops. This can be explained by noting that in these experiments \cite{RusBPJ2011,DrePNAS2013}, because of the flow passage geometry the streamer jet is not aligned to the background flow; rather, the jet can be assumed to be partly in a crossflow scenario (with respect to the background flow) (see Fig. 3 in Ref. \cite{Supp}). We show that in case the streamer jet is assumed to be completely in crossflow with the background flow, the breakup length can be even smaller than $R_s$ (see Ref. \cite{Supp} and Fig. 4 in Ref. \cite{Supp}) -- therefore the drop formation (from streamers) is caused by the presence of geometry-induced crossflow elements \cite{RusBPJ2011,DrePNAS2013}. Presence of such drops and its corresponding growth owing to the mass addition (see below) ensures the ``porous-matrix-like" structure inside the flow domain \cite{DrePNAS2013}. 
\begin{figure}[t!]
\centering
\includegraphics[width=9cm, height=4.6cm]{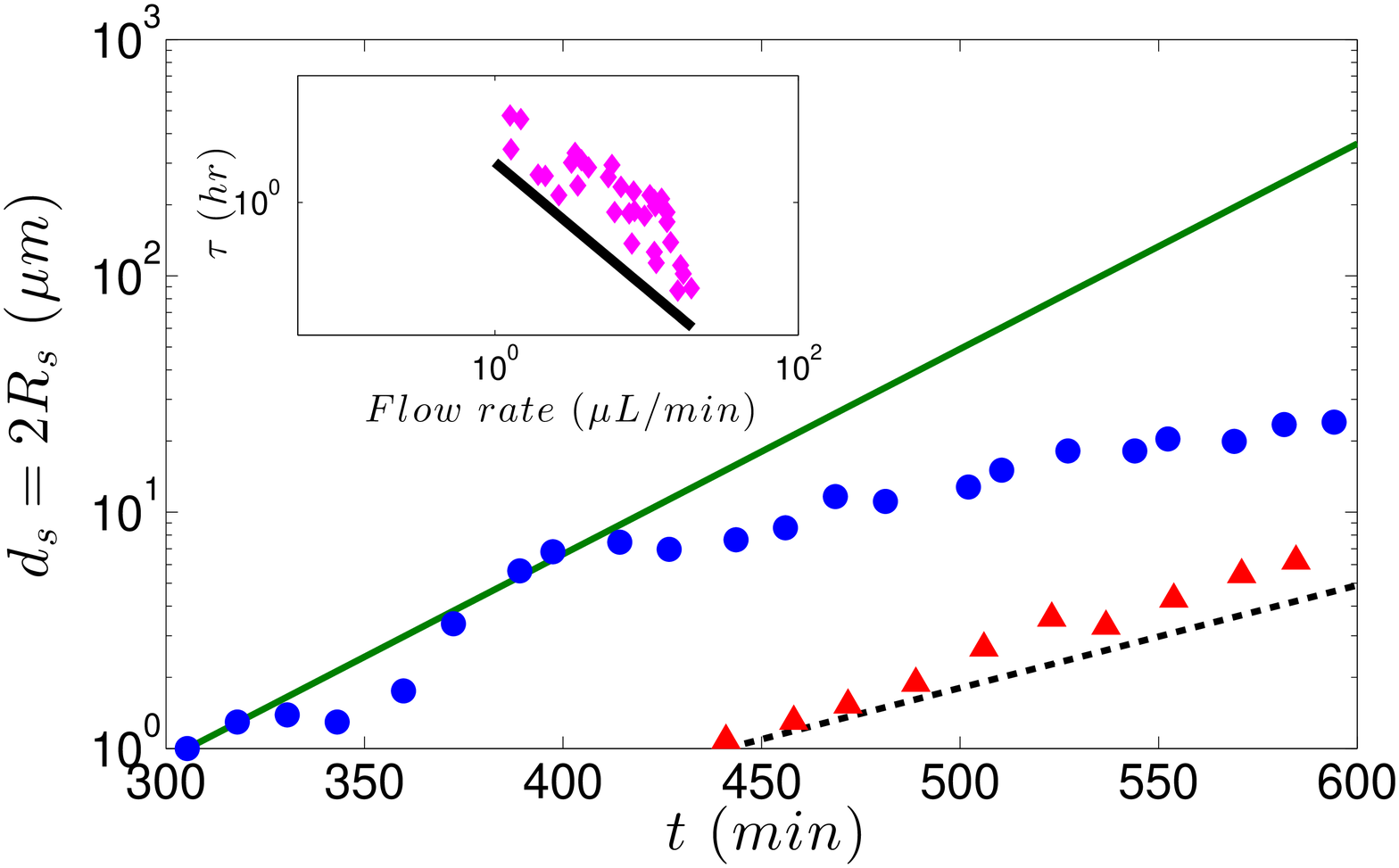}
\caption{(Color Online) Temporal variation of the streamer diameter $d_s=2R_s$. Experimental results correspond to the streamer formation in a microchannel with $330^\circ$ bend reported in the experiment of Rusconi et al. \cite{RusBPJ2011} for different flow rates (circles for $1~\mu L/min$ and triangles for $0.5~\mu L/min$). The continuous lines (solid line for flow rate for $1~\mu L/min$ and dashed line for $0.5~\mu L/min$) are the theoretical predictions using an exponential variation expressed as $R_s/R_{s,0}=d_s/d_{s,0}=\exp{\left[(t-t_0)/\tau_{theory,2}\right]}$ [this is a modification of eq.(\ref{eq:Rs_new}), considering $d_s(t=t_0)=d_{s,0}$; here $d_{s,0}$ and $t_0$ are the values obtained from the experiment]. To compute $\tau_{theory,2}$, we use eq.(\ref{eq:Rs_new}), with parameters $u_c=10^{-3}~m/s$ (for flow rate $1~\mu L/min$) and $u_c=0.5\times10^{-3}~m/s$ (for flow rate $0.5~\mu L/min$), $C=5\times10^{-4}cells/\mu m^3$ \cite{RusBPJ2011}, $A_{ac}=2\mu m^2$ and $\beta=0.67\times10^{-3}$. In the inset we show the variation of time scale ($\tau$) of streamer dynamics with the flow rate. Experimental results \cite{DrePNAS2013} are shown by markers, whereas our prediction ($\tau_{theory,2}$) is shown by a solid line. For this prediction, we consider the same dependence of $u_c$ on flow rate and keep all other parameters same, except for $C$, which is now $C=2\times10^{-4}cells/\mu m^3$ \cite{DrePNAS2013}.}
\label{fig:St_Th}       
\end{figure}    

\textit{Time variation of streamer dimensions}: 
The key physics behind the temporal variation of the streamer dimensions, as explained by Drescher et al. \cite{DrePNAS2013}, is the addition of mass to the streamers by the incoming cells. In their experiments, Rusconi et al. \cite{RusBPJ2011} reported a close to exponential (for small times) increase in the streamer dimensions with time. They also reported a smaller streamer dimension at a given time for a weaker flow rate. Drescher et al. \cite{DrePNAS2013} demonstrated that the time scale for this exponential streamer growth varies as $1/(flow~rate)$. We show \cite{Supp} that the analysis of Drescher et al. \cite{DrePNAS2013}, considering streamers as a ``solid" body, cannot recover this exponential behaviour -- on the contrary, their analysis will exhibit a growth dynamics as expressed in eq.(\ref{eq:Dre}) \cite{Supp}. More importantly, we demonstrate that such exponential increase in the streamer thickness can only be accounted by considering that the streamers evolve as drops \cite{Supp} [see eq.(\ref{eq:Rs_new}) for the corresponding growth dynamics]. 
%
\begin{equation}
\frac{R_s}{R_{s,0}}=\frac{1}{1-t/\tau_{theory,1}},~~\tau_{theory,1}=\frac{7\mu_f}{\beta C\Delta p A_{ac}R_{s,0}}.
\label{eq:Dre}
\end{equation}  
\begin{equation}
\frac{R_s}{R_{s,0}}=\exp{\left(t/\tau_{theory,2}\right)},~~\tau_{theroy,2}=\frac{2}{\beta Cu_cA_{ac}}.
\label{eq:Rs_new}
\end{equation}    
In the above equations, $C$ is the bacterial cell concentration, $A_{ac}$ is the area added by an advected cell to the streamer, $\beta$ is the fraction of cells that get caught in the streamer, $\Delta p$ is the pressure drop across the streamer (assuming it to be a ``solid" cylinder), and $R_{s,0}$ is the thickness of the streamer at the time when when streamer starts to form. In Fig.~\ref{fig:St_Th}, we show the comparison of the experimental results \cite{RusBPJ2011} and our theoretical prediction of the temporal variation of the streamer thickness for different flow rates. For smaller flow rates ($\sim0.5~\mu L/min$), we get excellent agreement with the experimental results, whereas for higher flow rates ($\sim1~\mu L/min$) the agreement is primarily at smaller times. At larger flow rates and at substantially large times, the flow clogging mechanism induced by the streamers will cause a weaker than exponential increase of the streamer dimensions. In the inset of the Fig.~\ref{fig:St_Th}, we compare our theoretical prediction ($\tau_{theory,2}$) with the experimental result \cite{DrePNAS2013} of the streamer formation time scale $\tau$ as a function of the flow rate, and recover the $1/(flow~rate)$ dependence of the time scale. Such dependence is also recovered for the time scale ($\tau_{theory,1}$) corresponding to the ``solid state" streamer, although the magnitude of $\tau_{theory,1}$ is substantially larger \cite{Supp}, which will fail to recover the $\tau$ values observed in experiments (see inset of Fig.~\ref{fig:St_Th}). 
Contrary to the studies of Rusconi et al. \cite{RusBPJ2011} and Drescher et al. \cite{DrePNAS2013}, in experimental set up of Valiei et al. \cite{ValLOC2012} the streamers being coaxial to the background flow, the flow direction is tangential to the direction of axis of the cylindrical jet (which does not break into droplets, see above), and hence the transfer of cells to the streamer can only occur diffusively (with diffusivity $D$), yielding \cite{Supp}
\begin{equation}
\frac{R_s}{R_{s,0}}=1+\frac{t}{\tau_{theory,3}},~~\tau_{theroy,3}=\frac{2R_{s,0}}{\beta CA_{ac}D}. 
\end{equation}  
Therefore, the streamer radius increases only linearly with time, which can explain the substantially weaker increase in the streamer thickness \cite{ValLOC2012}. 

\textit{Clogging effect of streamers}: As discussed by Drescher et al. \cite{DrePNAS2013}, one of the key signature of the streamer dynamics is the manner in which it clogs the flow by causing a substantial reduction in the flow rate. Drescher et al. \cite{DrePNAS2013} argued that such a behaviour can be attributed to the ``solid" state of the streamers and the fact that the streamers are positioned in the bulk and not at the walls. We show, on the contrary, that on being located in the bulk, the ``liquid" state of the streamers may actually lead to a greater reduction in the flow rate \cite{Supp} for certain ranges of streamer thickness values. This reduction gets severely more enhanced, and is manifested over the complete spectrum of the streamer thickness values, when the ``viscous liquid" streamer jet, on account of geometry-induced crossflow, break down into smaller dimensions, which would now occupy a much larger cross sectional area of the channel (see Ref. \cite{Supp} and Figs. 5 and 6 in Ref. \cite{Supp}).

To summarize, we have provided a theory to establish that the biofilm streamers, witnessed at very low Reynolds number ($Re\ll1$) microfluidic transport, form as viscous liquid jets. Our theory allows us to explain the very large time scales ($\sim$ several hours) associated with the streamer formation, that occurs in presence of extremely weak flow-driven shear stresses. Further, our theory reproduces the experimental results \cite{RusBPJ2011,DrePNAS2013} of growth dynamics of the streamers quantitatively, hitherto missing from the existing studies \cite{Exp_validation}. 
Finally, it is important to note that the streamer jets, which will invariably form as viscous jets, may attain viscoelastic rheology on account of entrapment of bacterial cells that produce EPS. The time scale of this change of rheology will be similar to that of the growth of process ($\sim hours$). Hence it will not affect the initial streamer viscous jet dynamics with much smaller characteristic time -- we probe this initial dynamics to explain the unbroken streamers in Valiei et al. \cite{ValLOC2012} and jet-to-drop transition in Rusconi et al. \cite{RusBPJ2011} and Drescher et al. \cite{DrePNAS2013}. But at larger times, this change of rheology can help explain issues such as the $C^{0.6}$ dependence of timescale dictating the streamer growth \cite{DrePNAS2013}, or the physical origin of the fitting factor $\beta$. Such explanation, along with those forwarded in this study will lead to a more comprehensive understanding of the biofilms in low Reynolds number hydrodynamics.

\end{document}